\documentclass[%
 reprint,
 amsmath,amssymb,
 aps,
 prl
]{revtex4-1}
\usepackage{graphicx}
\usepackage{dcolumn}
\usepackage{bm}
\usepackage{color}
\usepackage[dvipdfmx]{hyperref}

\begin{document}
\title{Flat-band Majorana bound states in topological Josephson junctions}
\author{Daisuke Oshima$^{1}$, Satoshi Ikegaya$^{1}$, Andreas P. Schnyder$^{2}$, and Yukio Tanaka$^{1}$}
\affiliation{$^{1}$Department of Applied Physics, Nagoya University, Nagoya 464-8603, Japan\\
$^{2}$Max-Planck-Institut f\"ur Festk\"orperforschung, Heisenbergstrasse 1, D-70569 Stuttgart, Germany}
\date{\today}
%
\begin{abstract}
Nodal topological superconductors characterized by $p_x$-wave pairing symmetry host flat-band Majorana bound states
causing drastic anomalies in low-energy electromagnetic responses.
Nevertheless, the study of flat-band Majorana bound states has been at a standstill owing to a serious lack of candidate materials for $p_x$-wave superconductors.
In this paper, by expanding a scheme of planar topological Josephson junctions,
we propose a promising device realizing an effective $p_x$-wave superconductor.
Specifically, we consider a three-dimensional Josephson junction
consisting of a thin-film semiconductor hosting a persistent spin-helix state and  two conventional $s$-wave superconductors.
We analytically obtain a topological phase diagram and numerically demonstrate the emergence of flat-band Majorana bound states by calculating the local density of states.
\end{abstract}

\maketitle
\section{Introduction}
Majorana bound states (MBSs) in topological superconductors (SCs) have been a central research topic in the field of superconductivity
\cite{kane_10,zhang_11,tanaka_12r,sato_17}.
An important finding of the intensive research efforts over the past decade is that MBSs have various forms
depending on the pairing symmetries of the parent topological SCs.
In two dimensions, for instance,
a chiral $p$-wave SC with broken time-reversal symmetry (TRS) hosts a MBS moving along the edge of the sample in one direction
(i.e., chiral MBS)~\cite{volovik_97,green_00,furusaki_01},
a helical $p$-wave SC preserving TRS harbors a pair of MBSs moving in opposite directions
(i.e., helical MBSs)~\cite{maiti_06,schnyder_08,zhang_09,tanaka_09},
and a nodal $p_x$-wave SC hosts flat-band MBSs~\cite{buchholtz_81,nagai_86,jerome_80,sarma_01,tanuma_01,sato_11}, will be the focus of this Letter.
A remarkable feature of the flat-band MBSs is that they have an extensive degeneracy at the Fermi level, unlike other forms of MBSs.
Accordingly, the flat-band MBSs of the $p_x$-wave SC affect low-energy electromagnetic responses drastically and cause various anomalies:
a prominent zero-bias conductance peak in normal-metal/SC junctions
\cite{bruder_90,hu_94,tanaka_95,tanaka_96,asano_04,tanaka_04,tanaka_05(1),asano_07,ikegaya_15,ikegaya_16(1)},
a fractional current-phase relationship in two-dimensional Josephson junctions~\cite{tanaka_96-(2),barash_96,kwon_04,asano_06(1),asano_06(2),ikegaya_16(2)},
and a distinct paramagnetic Meissner response to magnetic fields~\cite{higashitani_97,suzuki_14,suzuki_15}.

A crucial problem in the research of flat-band MBSs is a serious lack of candidate materials for $p_x$-wave SCs.
To resolve this problem, several theoretical models exhibiting effective $p_x$-wave superconductivity have been proposed, e.g., 
semiconductor/SC heterostructures~\cite{alicea_10,you_13,ikegaya_21,ikegaya_18}
and helical $p$-wave SCs under magnetic fields~\cite{ikegaya_18,law_13,rosenow_14}.
However, these proposals involve conditions that are difficult to achieve in experiments,
such as the necessity of a high Zeeman potential exceeding the pair potential of the parent SC and/or the absence of Rashba-type spin-orbit coupling (SOC) potentials,
while these systems intrinsically have structural inversion asymmetry causing a finite Rashba SOC~\cite{ikegaya_15,law_13}.

\begin{figure}[bbbb]
\begin{center}
\includegraphics[width=0.45\textwidth]{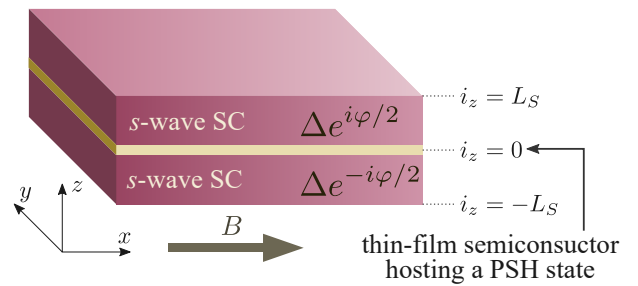}
\caption{Schematic of a Josephson junction consisting of two conventional $s$-wave SCs and a thin-film semiconductor hosting a PSH state.
 A magnetic field $B$ is applied in the $x$-direction.}
\label{fig:figure1}
\end{center}
\end{figure}

In this Letter, to resolve the stalemate in this research field, we propose an alternative route for creating an effective $p_x$-wave superconductor.
We specifically consider a three-dimensional Josephson junction illustrated in Fig.~\ref{fig:figure1},
where a thin-film semiconductor is sandwiched by two conventional $s$-wave SCs.
In addition, we assume that the semiconductor hosts a persistent spin-helix (PSH) state, which is studied intensively in the field of spintronics
\cite{bernevig_06,chang_06,schliemann_17,kohda_17}.
This work is inspired by the innovative studies on planar topological Josephson junctions (TJJs), which is attracting much attention recently
\cite{flensberg_17,halperin_17,haim_19,setiawan_19,ikegaya_20(2),shabani_20,yin_21,nichele_19,yacoby_19,shabani_21}.
In a planar TJJ, effective topological superconductivity is realized at the one-dimensional interfacial region of the two-dimensional Josephson junction.
In our setup, we obtain an effective $p_x$-wave superconductor in the vicinity of the two-dimensional semiconductor segment of the junction.
Thanks to the nature of TJJs~\cite{flensberg_17,halperin_17}, our setup can enter the topologically nontrivial phase supporting the flat-band MBSs
by applying Zeeman potentials smaller than the pair potentials in the superconducting segment.
Moreover, in the vicinity of the semiconductor segment, the local inversion symmetry along the $z$-direction is preserved, and thus the Rashba SOC can be negligible.
Therefore, our proposed system is free from the two difficult conditions required in previous proposals.
In addition, it has been recently shown that the PSH can be realized not only in III--V semiconductor compounds~\cite{awschalom_09,salis_12,kohda_12,salis_14},
but also in two-dimensional ferroelectric materials such as group-IV monochalcogenide monolayers
\cite{tsymbal_18,ishii_19,ishii_19(2),ishii_19(2),picozzi_19,zhao_19,jin_20,nardelli_20,chang_20}.
Hence, this rapid progress in the field of PSH makes it very likely that our proposal can be successfully fabricated in experiments.
Consequently, we propose a promising recipe for making an effective $p_x$-wave SC harboring flat-band MBSs.

\section{Effective Hamiltonian and Topological Phase Diagram}
We first derive an effective Hamiltonian that describes the essential properties of the proposed system.
Let us start with a simple Bogoliubov--de Gennes (BdG) Hamiltonian:
\begin{align}
\begin{split}
&\check{H}(\boldsymbol{r}) = \check{H}_0(\boldsymbol{r}) + \check{H}_{\lambda}(\boldsymbol{r}),\\
&\check{H}_0(\boldsymbol{r}) = \left[ \begin{array}{cccc}
\xi(\boldsymbol{r}) + V \hat{\sigma}_x & \Delta(z) (i\hat{\sigma}_y) \\
-\Delta^{\ast}(z) (i\hat{\sigma}_y) & -\xi(\boldsymbol{r}) \hat{\sigma}_0 - V \hat{\sigma}_x \\
\end{array} \right],\\
&\check{H}_{\lambda}(\boldsymbol{r}) = \left[ \begin{array}{cc}
-i\lambda \delta(z)\partial_x \hat{\sigma}_z & 0  \\
0  & -i\lambda \delta(z)\partial_x \hat{\sigma}_z \\
\end{array} \right],\\
&\xi(\boldsymbol{r}) = -\frac{\hbar^2}{2m} (\partial^2_x+\partial^2_y+\partial^2_z)-\mu,\\
&\Delta(z) = \left\{ \begin{array}{cl}
\Delta e^{i\varphi/2} & \text{for } z \leq 0\\
\Delta e^{-i\varphi/2} & \text{for } z > 0
\end{array} \right.,
\end{split}
\end{align}
where $m$, $\mu$, and $V$ represent the effective electron mass, the chemical potential,
and the Zeeman potential induced by the magnetic field along the $x$-direction, respectively.
The superconducting phase difference is given by $\varphi$.
The Pauli matrices in spin space are denoted by $\hat{\boldsymbol{\sigma}}=(\hat{\sigma}_x,\hat{\sigma}_y,\hat{\sigma}_z)$.
The junction interface is located at $z=0$, and we apply periodic boundary conditions in the $x$- and $y$-directions.
In this Hamiltonian, we do not describe explicitly the two-dimensional semiconductor hosting a PSH state.
Instead, we introduce $H_{\lambda}$
describing a unidirectional SOC potential generating the PSH~\cite{bernevig_06,chang_06,schliemann_17,kohda_17} only at $z=0$.
The SOC potential as in $H_{\lambda}$ is realized in
zinc-blende III--V semiconductor quantum wells grown along the $[110]$ direction~\cite{bernevig_06,chang_06,salis_14},
in quantum wells with equal amplitudes of Rashba and Dresselhaus SOC~\cite{bernevig_06,chang_06,awschalom_09,salis_12,kohda_12},
and in ferroelectric thin-film materials~\cite{tsymbal_18,ishii_19,ishii_19(2),picozzi_19,zhao_19,jin_20,nardelli_20,chang_20}.
In what follows, we construct an effective low-energy Hamiltonian that focuses on the subgap states localized in the vicinity of the junction interface.
To simplify the discussion, we treat $H_{\lambda}$ as a perturbation.
Within zeroth-order perturbation, we first solve the BdG equation for the subgap states,
\begin{align}
\begin{split}
&\int d \boldsymbol{r} \check{U}^{\dagger}_{\mathrm{SG}}(\boldsymbol{r}) \check{H}_0(\boldsymbol{r}) \check{U}_{\mathrm{SG}}(\boldsymbol{r})
= \check{E}_0,\\
&\check{E}_0=\mathrm{diag}[\varepsilon-V, \varepsilon+V, -\varepsilon+V, -\varepsilon-V],\\
&\check{U}_{\mathrm{SG}}(\boldsymbol{r}) = \check{U}_{\mathrm{SG}}(z) \frac{e^{ik_x x} e^{ik_y y}}{\sqrt{L_x L_y}},
\end{split}
\end{align}
where the conditions,
\begin{align}
0 < \varepsilon < \Delta, \quad
\check{U}_{\mathrm{SG}}(z=\pm\infty) = 0,
\end{align}
are satisfied.
Here $k_{x(y)}$ and $L_{x(y)}$ represent the momentum and width in the $x$-direction ($y$-direction), respectively.
The explicit form of $\check{U}_{\mathrm{SG}}(\boldsymbol{r})$ is given in the Supplemental Material (SM)~\cite{sup_mat}.
Then, on the basis of the first-order perturbation theory, we construct an effective Hamiltonian for the subgap states,
\begin{align}
\begin{split}
\check{H}_{\mathrm{eff}} &= \int d \boldsymbol{r}
\check{U}^{\dagger}_{\mathrm{SG}}(\boldsymbol{r}) \check{H}(\boldsymbol{r}) \check{U}_{\mathrm{SG}}(\boldsymbol{r}) \\
&=\left[ \begin{array}{cccc}
\varepsilon -V & \tilde{\lambda}k_x & \tilde{\Delta}k_x & 0 \\
\tilde{\lambda}k_x & \varepsilon +V & 0 & -\tilde{\Delta}k_x \\ 
\tilde{\Delta}k_x & 0 & -\varepsilon +V & \tilde{\lambda}k_x \\
0 & -\tilde{\Delta}k_x & \tilde{\lambda}k_x & -\varepsilon -V \\
\end{array} \right],
\label{eq:eff_ham}
\end{split}
\end{align}
with
$\varepsilon = \Delta \sqrt{1-D \sin^2 (\varphi/2)}$,
$\tilde{\lambda} = \lambda \sqrt{D(1-D)} F$, and
$\tilde{\Delta} = \lambda \cot (\varphi/2) F$,
where $D=D(k_x,k_y,\varphi)$ and $F=F(k_x,k_y,\varphi)$ are the even functions of $k_x$, $k_y$, and $\varphi$.
The explicit forms of $D(k_x,k_y,\varphi)$ and $F(k_x,k_y,\varphi)$ are given in the SM~\cite{sup_mat}.
The effective Hamiltonian of $\check{H}_{\mathrm{eff}}$ is equivalent to the BdG Hamiltonian of a two-dimensional $p_x$-wave SC
with SOC and Zeeman potential.
Namely, we can reinterpret
$\varepsilon$ as the kinetic energy of the quasi-particle states confined in the two-dimensional space
(i.e., the subgap states localized at the junction interface),
$\tilde{\lambda}$ as the effective SOC potential,
and most importantly, $\tilde{\Delta}$ as the effective $p_x$-wave pair potential acting on the quasi-particle states.
Consequently, we can expect that the present junction hosts flat-band MBSs at the surfaces perpendicular to the $x$-direction.

\begin{figure}[hhhh]
\begin{center}
\includegraphics[width=0.5\textwidth]{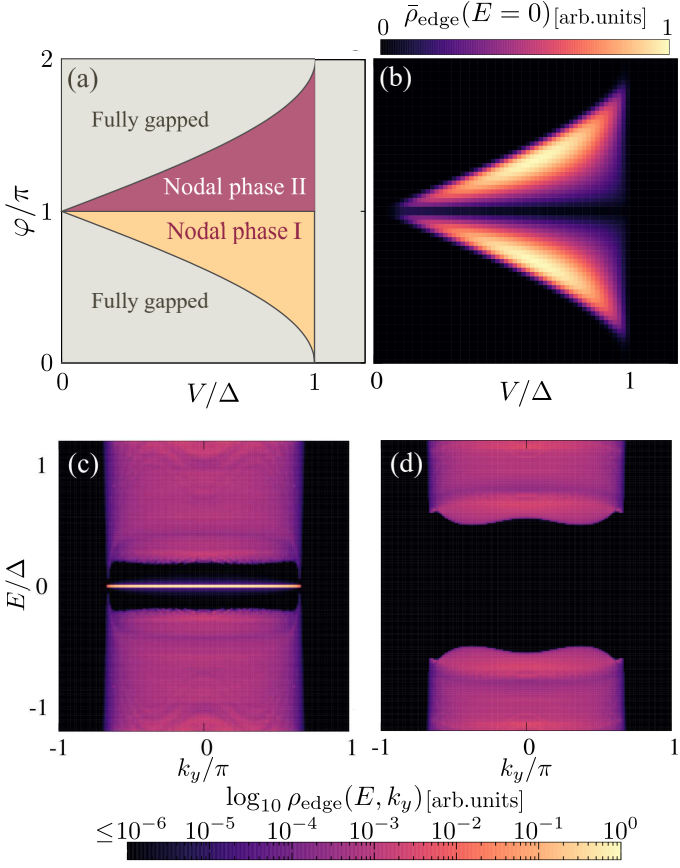}
\caption{(a) Phase diagram of the nodal structure.
The nodal phases satisfy the condition $\Delta^2 \cos (\varphi/2) < V^2 <\Delta^2$, which is obtained in the limit of $\mu \gg \Delta$.
The nodal phase I and II are characterized by the positive and negative winding number, respectively.
Simultaneously, the both nodal phases are characterized by the nontrivial $\mathbb{Z}_2$ topological number.
(b) Zero-energy LDOS,  $\bar{\rho}_{\rm edge}(E=0)$, as a function of the Zeeman potential $V$ and the superconducting phase difference $\varphi$.
In (c) and (d), we show $\rho_{\rm edge}(k_y;E)$ on a log-scale, i.e., $\log_{10} \rho_{\rm edge}(k_y;E)$, as a function of the energy $E$ and the momentum $k_y$.
We plot the result for (c) the topologically nontrivial phase with $(V,\varphi)=(0.7\Delta,0.65\pi)$
and that for (d) the topologically trivial phase with $(V,\varphi)=(0.25\Delta,0.25\pi)$, respectively.}
\label{fig:figure2}
\end{center}
\end{figure}

To clarify the emergence of the flat-band MBSs, we analyze the topological property of $\check{H}_{\mathrm{eff}}$.
By diagonalizing $\check{H}_{\mathrm{eff}}$, we obtain the energy eigenvalues as
\begin{align}
E_{s}=\pm \sqrt{\left( \varepsilon + s \sqrt{V^2+\tilde{\lambda}^2k_x^2}\right)^2 + \tilde{\Delta}^2 k_x^2},
\end{align}
for $s=\pm$.
As discussed in the SM~\cite{sup_mat} in detail, we obtain the two gap nodes when the condition $X < V^2 <\Delta^2$ is satisfied,
where the lower bound $X$ is given explicitly in the SM and satisfies
\begin{align}
\lim_{\Delta/\mu \rightarrow 0} X = \Delta^2 \cos^2 (\varphi/2).
\end{align} 
The two gap nodes are located at $(k_x,k_y) = (0,\pm k_F \sqrt{1+\zeta})$ with
\begin{align}
\zeta = \frac{1}{\mu}\left[V^2-\frac{\Delta^2}{2}\left\{1+\cos^2 \frac{\varphi}{2}\right\}\right]\sqrt{\frac{1}
{V^2-\Delta^2 \cos^2 \frac{\varphi}{2}}}, \nonumber
\end{align}
where $k_F=\sqrt{2m \mu}/\hbar$.
Moreover, the energy spectrum becomes completely gapless at $\varphi=\pi$, because the effective pair potential $\tilde{\Delta} \propto \cot(\phi/2)$ vanishes.
Consequently, we obtain two distinct nodal superconducting phases as summarized in Fig.~\ref{fig:figure2}(a):
the nodal phase I for $\varphi<\pi$ and the nodal phase II for $\varphi>\pi$.

Next, we discuss the symmetry properties of the present junction. 
The effective Hamiltonian $\check{H}_{\mathrm{eff}}$ has TRS and particle-hole symmetry (PHS) as,
\begin{align}
&\check{T}_{\mathrm{eff}} \check{H}_{\mathrm{eff}}(\boldsymbol{k}) \check{T}_{\mathrm{eff}}^{-1}
=\check{H}_{\mathrm{eff}}(-\boldsymbol{k}),\quad
\check{T}_{\mathrm{eff}}= -i \hat{\sigma}_z \check{\tau}_z \mathcal{K},\\
&\check{C}_{\mathrm{eff}} \check{H}_{\mathrm{eff}}(\boldsymbol{k}) \check{C}_{\mathrm{eff}}^{-1}
=-\check{H}_{\mathrm{eff}}(-\boldsymbol{k}),\quad
\check{C}_{\mathrm{eff}}= \check{\tau}_x \mathcal{K},
\end{align}
where the Pauli matrices in Nambu space are given by $\check{\boldsymbol{\tau}}=(\check{\tau}_x,\check{\tau}_y,\check{\tau}_z)$,
and $\mathcal{K}$ is the complex-conjugation operator.
By combining $\check{T}_{\mathrm{eff}}$ and $\check{C}_{\mathrm{eff}}$, we obtain a chiral symmetry (CS) as
\begin{align}
\check{S}_{\mathrm{eff}} \check{H}_{\mathrm{eff}}(\boldsymbol{k}) \check{S}_{\mathrm{eff}}^{-1}
=-\check{H}_{\mathrm{eff}}(\boldsymbol{k}),\quad
\check{S}_{\mathrm{eff}}=\check{T}_{\mathrm{eff}}\check{C}_{\mathrm{eff}}.
\end{align}
The original Hamiltonian $\check{H}(\boldsymbol{r})$ also preserves TRS, PHS, and CS as
\begin{align}
&\check{T} \check{H} (\boldsymbol{r}) \check{T}^{-1}=\check{H} (\boldsymbol{r}),\quad
\check{T}=\check{M}_{xy}\check{T}_-,\\
&\check{C} \check{H} (\boldsymbol{r}) \check{C}^{-1}=-\check{H} (\boldsymbol{r}),\quad
\check{C} = \check{\tau}_x \mathcal{K},\\
&\check{S} \check{H} (\boldsymbol{r}) \check{S}^{-1}=-\check{H} (\boldsymbol{r}),\quad
\check{S}=\check{T}\check{C},
\label{eq:chiral_symm}
\end{align}
where $\check{M}_{xy}=-i \hat{\sigma}_z \check{\tau}_z \mathcal{R}_z$ denotes mirror reflection symmetry with respect to the $xy$-plane,
and $\check{T}_-=i \hat{\sigma}_y \mathcal{K}$ represents conventional TRS obeying $\check{T}_-^2=-1$;
$\mathcal{R}_z$ describes the reflection of the spatial coordinate $z$ (i.e., $z \rightarrow -z$).
Furthermore, the present system preserves the inversion symmetry with respect to the $y$-axis:
\begin{align}
&\check{R}_{k_y} \check{H}_{\mathrm{eff}}(\boldsymbol{k}) \check{R}_{k_y}^{-1}=
\check{H}_{\mathrm{eff}}(k_x,-k_y)=\check{H}_{\mathrm{eff}}(\boldsymbol{k}), \\
&\check{R}_{y} \check{H}(\boldsymbol{r}) \check{R}_{y}^{-1} = \check{H}(x,-y,z)=\check{H}(\boldsymbol{r}),
\end{align}
where $\check{R}_{k_y}$ ($\check{R}_{y}$) inverts the sign of $k_y$ ($y$) as $k_y \rightarrow -k_y$ ($y \rightarrow -y$).
As a result, we additionally obtain a modified TRS and a modified PHS as
\begin{align}
\begin{split}
&\check{\underline{T}}_{\mathrm{eff}} \check{H}_{\mathrm{eff}}(\boldsymbol{k}) \check{\underline{T}}_{\mathrm{eff}}^{-1}
=\check{H}_{\mathrm{eff}}(-k_x,k_y), \\
&\check{\underline{T}}_{\mathrm{eff}}=\check{R}_{k_y}\check{T}_{\mathrm{eff}},
\end{split} \label{eq:mod_trs}
\end{align}
and
\begin{align}
\begin{split}
&\check{\underline{C}}_{\mathrm{eff}} \check{H}_{\mathrm{eff}}(\boldsymbol{k}) \check{\underline{C}}_{\mathrm{eff}}^{-1}
=-\check{H}_{\mathrm{eff}}(-k_x,k_y), \\
&\check{\underline{C}}_{\mathrm{eff}}=\check{R}_{k_y}\check{C}_{\mathrm{eff}},
\end{split} \label{eq:mod_phs}
\end{align}
where these modified symmetries do not involve the sign change in $k_y$.
The original BdG Hamiltonian also preserves
\begin{gather}
\check{\underline{T}} \check{H}(\boldsymbol{r}) \check{\underline{T}}^{-1}
=\check{H}(\boldsymbol{r}), \quad
\check{\underline{T}}=\check{R}_{y}\check{T},\\
\check{\underline{C}}\check{H}(\boldsymbol{r}) \check{\underline{C}}^{-1}
=-\check{H}(\boldsymbol{r}), \quad
\check{\underline{C}}=\check{R}_{y}\check{C}.\label{eq:mod_phs_original}
\end{gather}
The relevant CS operator satisfies
\begin{gather}
\check{\underline{S}}_{\mathrm{eff}}=\check{\underline{T}}_{\mathrm{eff}}\check{\underline{C}}_{\mathrm{eff}}=\check{S}_{\mathrm{eff}},\\
\check{\underline{S}}=\check{\underline{T}}\;\check{\underline{C}}=\check{S}.
\end{gather}
To characterize the nodal phases of $\check{H}_{\mathrm{eff}}(k_x,k_y)$ topologically, 
we set $k_y$ to a fixed value, and study $\check{H}_{\mathrm{eff}}$ as a function of only $k_x$.
I.e, we interpret $\check{H}_{\mathrm{eff}}(k_x,k_y)$ as a one parameter family of one-dimensional Hamiltonians
with momentum $k_x$ and parameter $k_y$~\cite{sato_11}.
Since $\check{H}_{\mathrm{eff}}(k_x,k_y)$ satisfies Eq.~(\ref{eq:mod_trs}) and Eq.~(\ref{eq:mod_phs}) with
$\check{\underline{T}}_{\mathrm{eff}}^2=+1$ and $\check{\underline{C}}_{\mathrm{eff}}^2=+1$,
we can classify $\check{H}_{\mathrm{eff}}(k_x,k_y)$ into the BDI symmetry class in one-dimension~\cite{schnyder_08}.
Therefore, the topological property of $\check{H}_{\mathrm{eff}}(k_x,k_y)$ can be characterized by a one-dimensional winding number~\cite{schnyder_08,sato_11},
\begin{align}
w(k_y) = \frac{i}{4 \pi}
\int dk_x \mathrm{Tr} [ \check{S}_{\mathrm{eff}} \check{H}_{\mathrm{eff}}^{-1}
\partial_{k_x} \check{H}_{\mathrm{eff}}].
\label{eq:wind_num}
\end{align}
By computing this as in the SM~\cite{sup_mat}, we obtain the winding number for the nodal phase I (phase II) as
\begin{align}
w(k_y)=
\left\{ \begin{array}{cl} 
+(-)1& \text{for}\quad |k_y|< k_F \sqrt{1+\zeta} \\
0 & \text{otherwise}
\end{array}\right.,
\end{align}
where $w(k_y)=0$ irrespective of $k_y$ for the fully gapped phase.
Since the winding number is nonzero in a finite range of $k_y$, according to the bulk-boundary correspondence,
we obtain flat-band MBSs at a surface perpendicular to the $x$-direction~\cite{sato_11}.
Unless the gap of $\tilde{\Delta}$ closes or the CS of $\check{S}_{\mathrm{eff}}$ (and therefore $\check{S}$) is broken,
the flat-band MBSs can exist stably due to the topological protection.
Therefore, we can expect that higher-order perturbations of $\lambda$ do not disturb the flat-band MBSs.

Moreover, we can additionally define a $\mathbb{Z}_2$ topological number as $\mathbb{Z}_2(k_y)=(-1)^{w(k_y)}$~\cite{halperin_17}.
In terms of the $\mathbb{Z}_2$ topological number, the flat-band MBSs in both nodal phases I and II  are characterized by $\mathbb{Z}_2(k_y)=-1$.
When some perturbation breaks the modified TRS and the CS,
the symmetry class of $\check{H}_{\mathrm{eff}}(k_x,k_y)$ changes into class D~\cite{schnyder_08}.
Nevertheless, as long as the modified PHS of $\check{\underline{C}}_{\mathrm{eff}}$ (and therefore $\check{\underline{C}}$) is preserved,
the flat-band MBSs still exist, but are now characterized by the $\mathbb{Z}_2$ topological number~\cite{schnyder_08,halperin_17}, rather than the winding number.
Hence, perturbations breaking the modified TRS might alter the topological phase boundaries, but do not remove the flat-band MBSs.
When some perturbation breaks the modified PHS and the modified TS, 
the symmetry class of $\check{H}_{\mathrm{eff}}(k_x,k_y)$ changes into class AIII~\cite{schnyder_08}.
In this case, the system still preserves CS of $\check{S}_{\mathrm{eff}}$ (and therefore $\check{S}$),
leading to flat-band MBSs characterized by the winding number $w(k_y)$.
In summary, we can expect the emergence of flat-band MBSs as long as the present junction preserves the CS or the modified PHS.

\section{Flat-band Majorana bound states}
To check the emergence of the flat-band MBSs, we describe the present junction using a tight-binding model
and calculate numerically the local density of states (LDOS) at the surface perpendicular to the $x$-direction.
We denote the lattice constant by $a_0$ and parameterize the lattice sites by $\bm{r}=i_x\bm{x}+i_y\bm{y}+i_z\bm{z}$, with $|\bm{x}|=|\bm{y}|=|\bm{z}|=a_0$.  
The upper (lower) SC occupies $1\leq i_z \leq L_S$ ($-L_S\leq i_z \leq -1$),
and the two-dimensional semiconductor is located at $i_z=0$ (see Fig.~\ref{fig:figure1}).
We apply a periodic (an open) boundary condition in the $y$-direction ($z$-direction).
We assume that the system is semi-infinite in the $x$-direction, i.e., $0\leq i_x \leq \infty$.
For latter convenience, we apply a Fourier transformation on the annihilation operator at $\bm{r}$ with spin $\sigma$ as
$c_{\bm{r},\sigma}=\sum_{k_y} e^{ik_y a_0 i_y} c_{\tilde{\bm{r}},k_y,\sigma}/\sqrt{L_y}$, where $\tilde{\bm{r}}=i_x\bm{x}+i_z\bm{z}$.
Furthermore, in what follows, we omit the subscript of $k_y$ in the annihilation operator for notational simplicity,
i.e., $c_{\tilde{\bm{r}},\sigma}$ means $c_{\tilde{\bm{r}},k_y,\sigma}$.
The BdG tight-binding Hamiltonian reads $H = \sum_{k_y} H(k_y)$ with
\begin{align}
H(k_y)=&-t\sum_{\tilde{\bm{r}}}\sum_{\sigma=\uparrow,\downarrow}
\left(c^\dagger_{\tilde{\bm{r}}+\bm{x},\sigma}c_{\tilde{\bm{r}},\sigma}+ c^\dagger_{\tilde{\bm{r}},\sigma}c_{\tilde{\bm{r}}+\bm{x},\sigma}\right)
\nonumber\\
& -t\sum_{\tilde{\bm{r}},\sigma}
\left(c^\dagger_{\tilde{\bm{r}}+\bm{z},\sigma}c_{\tilde{\bm{r}},\sigma}+ c^\dagger_{\tilde{\bm{r}},\sigma}c_{\tilde{\bm{r}}+\bm{z},\sigma}\right)
\nonumber\\
&+\sum_{\tilde{\bm{r}},\sigma} \xi_{i_z} c^{\dagger}_{\tilde{\bm{r}},\sigma}c_{\tilde{\bm{r}},\sigma}
+\sum_{\tilde{\bm{r}},\sigma,\sigma^{\prime}} V (\hat{\sigma}_y)_{\sigma,\sigma^{\prime}} c^{\dagger}_{\tilde{\bm{r}},\sigma}c_{\tilde{\bm{r}},\sigma^{\prime}}
\nonumber\\
&+\sum_{i_x}\sum_{i_z=1}^{L_S} \Delta e^{i\varphi/2} ( c^{\dagger}_{\tilde{\bm{r}},k_y,\uparrow}c^{\dagger}_{\tilde{\bm{r}},-k_y,\uparrow} + \mathrm{H.c.})
\nonumber\\
&+\sum_{i_x}\sum_{i_z=-L_S}^{-1} \Delta e^{-i\varphi/2} ( c^{\dagger}_{\tilde{\bm{r}},k_y,\uparrow}c^{\dagger}_{\tilde{\bm{r}},-k_y,\uparrow} + \mathrm{H.c.})
\nonumber\\
&+\sum_{\tilde{\bm{r}},\sigma,\sigma^{\prime}}\frac{i \lambda}{2} \delta_{i_z,0} (\hat{\sigma}_z)_{\sigma,\sigma^{\prime}} \nonumber\\
&\qquad \times \left(c^\dagger_{\tilde{\bm{r}}+\bm{x},\sigma}c_{\tilde{\bm{r}},\sigma^{\prime}}
- c^\dagger_{\tilde{\bm{r}},\sigma}c_{\tilde{\bm{r}}+\bm{x},\sigma^{\prime}}\right),
\label{eq:lat_ham}\\
\xi_{i_z} =&\left\{\begin{aligned}
4 t-\mu_N -2t \cos(k_y a_0) \quad &{\rm for} \quad i_z=0\\
6 t-\mu_S -2t \cos(k_y a_0) \quad &{\rm otherwise}
\end{aligned}\right.,\,
\end{align}
where $t$ is the nearest-neighbor hopping integral, and $\mu_{S(N)}$ is the chemical potential in the SC (semiconductor).
From the BdG Hamiltonian in Eq.~(\ref{eq:lat_ham}), we calculate the Green's function $G_{k_y}(\tilde{\bm{r}},\tilde{\bm{r}}^{\prime}; E)$
for each $k_y$ by using recursive Green's function techniques~\cite{fisher_81,ando_91}. 
The LDOS is evaluated by
\begin{align}
\rho_{k_y}(\tilde{\bm{r}};E)=-\frac{1}{\pi}\left\{{\rm Tr}\left[G_{k_y}(\tilde{\bm{r}},\tilde{\bm{r}};E+i\delta)\right]\right\},
\end{align}
where $\delta$ is a small positive value added to the energy $E$.
We specifically focus on the LDOS at the edge of the semiconductor segment (i.e., $i_x=i_z=0$):
\begin{align}
&\rho_{\rm edge}(k_y;E)=\rho_{k_y}(i_x=0,i_z=0;E),\\
&\bar{\rho}_{\rm edge}(E)=\sum_{k_{y}}\rho_{\rm edge}(k_y;E).
\end{align}
Hereafter, we fix the parameters as $t=10 \Delta$, $\mu_S=30 \Delta$, $\mu_N=10 \Delta$, $\lambda=10 \Delta$, $\delta=10^{-4}\Delta$ and $L_S=30$.
Figure~\ref{fig:figure2}(b) shows the zero-energy LDOS, $\bar{\rho}_{\rm edge}(E=0)$, as a function of $V$ and $\varphi$.
We see that $\bar{\rho}_{\rm edge}(E=0)$ increases significantly
in the parameter regions corresponding to the topologically nontrivial phases in Fig.~\ref{fig:figure2}(a).
In Figs.~\ref{fig:figure2}(c) and ~\ref{fig:figure2}(d),
we show $\rho_{\rm edge}(k_y;E)$ for (c) the topologically nontrivial phase with $(V,\varphi)=(0.7\Delta,0.65\pi)$
and for (d) the topologically trivial phase with $(V,\varphi)=(0.25\Delta,0.25\pi)$, respectively.
We plot $\log_{10} \rho_{\rm edge}(k_y;E)$ instead of the raw data of $\rho_{\rm edge}(k_y;E)$.
As shown in Fig.~\ref{fig:figure2}(c), the LDOS in the topologically nontrivial phase shows the drastic enhancement at $E=0$ for the broad range of $k_y$,
while that in the topologically trivial phase has a gapped structure.
On the basis of both analytical and numerical results, we demonstrate the emergence of the flat-band MBSs due to the effective $p_x$-wave superconductivity.

\begin{figure}[bbbb]
\begin{center}
\includegraphics[width=0.5\textwidth]{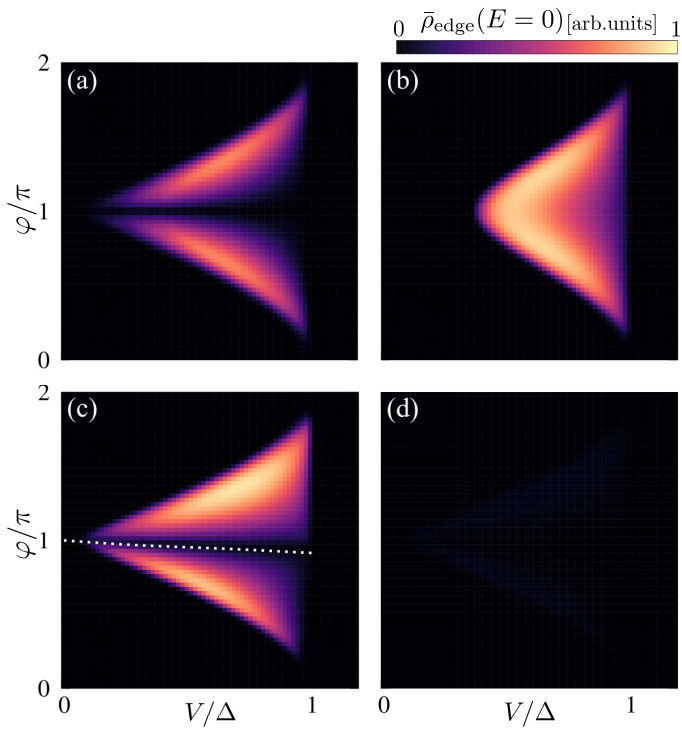}
\caption{The zero-energy LDOS, $\bar{\rho}_{\rm edge}(E=0)$, as a function of $V$ and $\varphi$.
The results are normalized by the same unit as the one in Fig.~\ref{fig:figure2}(b).
In (a) and (b), we consider (a) $(t_1,t_2)=(-2\Delta,-2\Delta)$ and (b) $(t_1,t_2)=(-2\Delta,2\Delta)$, respectively.
In (a),  both CS in Eq.~(\ref{eq:chiral_symm}) and the modified PHS in Eq.~(\ref{eq:mod_phs_original}) are preserved.
In (b), the modified PHS is preserved, while the CS is broken.
In (c) and (d), we introduce the Rashba SOC potential with $(\lambda^{\prime}_1,\lambda^{\prime}_2)=(2.5\Delta,-2.5\Delta)$ 
and (d) $(\lambda^{\prime}_1,\lambda^{\prime}_2)=(3.5\Delta,-1.5\Delta)$, respectively.
In (c), the CS is preserved, while the modified PHS is broken.
In (d), both CS and modified PHS are broken.
The dotted line in (c) indicates $(V,\varphi)$ taking the local minima of $\bar{\rho}_{\rm edge}(E=0)$.}
\label{fig:figure3}
\end{center}
\end{figure}
Finally, we discuss the stability of the flat-band MBSs against perturbations.
We first consider a perturbation of the form,
\begin{align}
H_{\delta t} = -\sum_{\tilde{\bm{r}},\sigma} &(t_1 \delta_{i_z,1} + t_2 \delta_{i_z,-1}) \nonumber\\
& \times \left(c^\dagger_{\tilde{\bm{r}}+\bm{z},\sigma}c_{\tilde{\bm{r}},\sigma}+ c^\dagger_{\tilde{\bm{r}},\sigma}c_{\tilde{\bm{r}}+\bm{z},\sigma}\right),
\end{align}
which lowers the transparency at the superconductor/semiconductor interfaces.
When $t_1=t_2$, the system preserves both CS in Eq.~(\ref{eq:chiral_symm}) and modified PHS in Eq.~(\ref{eq:mod_phs_original}).
The figure~\ref{fig:figure3}(a) shows $\bar{\rho}_{\rm edge}(E=0)$ in the chiral symmetric case as a function of $V$ and $\varphi$, where we choose $t_1=t_2=-2\Delta$.
Since the CS and the modified PHS are preserved, the flat-band MBSs can remain at zero-energy, and therefore we find a significant enhancement in the LDOS.
Even so, as compared with Fig.~\ref{fig:figure2}(b),  we find that the LDOS around $(V,\varphi) = (0,\pi)$ is suppressed.
Furthermore, by calculating $\rho_{\rm edge}(k_y;E)$, we see that
the induced gap (which corresponds to $\tilde{\Delta}$ in the first-order perturbation theory) is suppressed by nonzero $t_{1(2)}$ (see the SM~\cite{sup_mat}).
The shrinking of the topologically nontrivial phase and the suppression of the induced gap owing to the reduction of the junction transparency
are also found in the planar TJJs~\cite{halperin_17}.
When $t_1\neq t_2$, the system preserves the modified PHS, while the CS is broken.
In Fig.~\ref{fig:figure3}(b), we show $\bar{\rho}_{\rm edge}(E=0)$, where we choose $t_1=-2\Delta$ and $t_2=2\Delta$.
Even in the absence of the CS, since the modified PHS is preserved, we still find a significant enhancement in the LDOS.
By calculating $\rho_{\rm edge}(k_y;E)$, we also confirm that the enhancement is indeed caused by the flat-band MBSs (see also SM~\cite{sup_mat}).
We note that the flat-band MBSs with $t_1 \neq t_2$ are no longer characterized by the winding number in Eq.~(\ref{eq:wind_num}),
but by the nontrivial $\mathbb{Z}_2$ topological number.
We additionally find that the two nodal phases originally separated by $\varphi=\pi$ seem to merge into a single nodal phase.
Then, we consider another perturbation,
\begin{align}
H_{R} =& \sum_{\tilde{\bm{r}},\sigma,\sigma^{\prime}}
\left\{ \lambda^{\prime}_1 \sin(k_y a_0) \delta_{i_z,1} + \lambda^{\prime}_2 \sin(k_y a_0) \delta_{i_z,-1} \right\} \nonumber\\
 & \qquad \times (\hat{\sigma}_x)_{\sigma,\sigma^{\prime}} c^{\dagger}_{\tilde{\bm{r}},\sigma}c_{\tilde{\bm{r}},\sigma^{\prime}} \nonumber\\
-& \sum_{\tilde{\bm{r}},\sigma,\sigma^{\prime}}
\left\{ \frac{i \lambda^{\prime}_1}{2}\delta_{i_z,1} + \frac{i \lambda^{\prime}_2}{2} \delta_{i_z,-1} \right\}\nonumber\\
&\qquad \times (\hat{\sigma}_y)_{\sigma,\sigma^{\prime}}  \left(c^\dagger_{\tilde{\bm{r}}+\bm{x},\sigma}c_{\tilde{\bm{r}},\sigma^{\prime}}
- c^\dagger_{\tilde{\bm{r}},\sigma}c_{\tilde{\bm{r}}+\bm{x},\sigma^{\prime}}\right),
\end{align}
which describes the Rashba SOC potentials at one layer above and below the semiconductor segment.
When $\lambda^{\prime}_1=-\lambda^{\prime}_2$, the junction preserves the CS, while the modified PHS is broken.
In Fig.~\ref{fig:figure3}(c), we plot $\bar{\rho}_{\rm edge}(E=0)$, where we choose $\lambda^{\prime}_1=-\lambda^{\prime}_2=2.5\Delta$.
Since the CS is preserved, we still find a significant enhancement in the LDOS due to the flat-band MBSs.
We also find that $\bar{\rho}_{\rm edge}(E=0)$ becomes asymmetric with respect to $\varphi=\pi$, i.e., the topologically nontrivial region for $\varphi < \pi$ shrinks.
By calculating $\rho_{\rm edge}(k_y;E)$, we also confirm that the amplitude of the induced gap also becomes asymmetric with respect to $\varphi=\pi$,
i.e., the induced gap for $\varphi < \pi$ is smaller than that for $\varphi > \pi$ (see SM~\cite{sup_mat}).
Strictly speaking, the phase boundary between two distinct nodal phases slightly deviates from $\varphi=\pi$; see dotted line in Fig.~\ref{fig:figure3}(c).
When $\lambda^{\prime}_1\neq-\lambda^{\prime}_2$, both CS and modified PHS are broken.
In Fig.~\ref{fig:figure3}(d), we show $\bar{\rho}_{\rm edge}(E=0)$ with $\lambda^{\prime}_1=3.5\Delta$ and $\lambda^{\prime}_2=-1.5\Delta$.
We see that the zero-energy LDOS is strongly suppressed, which implies the absence of the flat-band MBSs.
More specifically, by calculating $\rho_{\rm edge}(k_y;E)$,
we confirm that the Rashba SOC potential causes the finite slope in the dispersion of flat-band MBSs (see SM~\cite{sup_mat}).
The flat-band MBSs in other effective $p_x$-wave superconductors are also destroyed by the Rashba SOC potential in a similar manner~\cite{ikegaya_15,law_13}.
However, we emphasize that the present junction intrinsically preserves inversion symmetry with respect to the $z$-direction,
and thus the Rashba SOC potential with $\lambda^{\prime}_1\neq -\lambda^{\prime}_2$ becomes negligibly small.
In short, although the perturbations shrink the parameter regions for the topologically nontrivial phase,
we can obtain the flat-band MBSs as long as the Josephson junction preserves the CS or the modified PHS.

\section{Discussion}
Within our numerical calculations,
we obtain a sizable induced gap corresponding to almost $25\%$ of the parent pair potential $\Delta$,
which is comparable to the induced gap of the planar TJJ~\cite{halperin_17}.
When we employ Nb ($\Delta \approx 1.5 \mathrm{meV}$~\cite{sutton_62}) for the superconducting segment,
for instance, the induced gap is up to $ 375 \mu \mathrm{eV}$.
The induced gap becomes larger with a larger SOC potential sustaining the PSH state and with higher transparency at the semiconductor/superconductor interface.
This property of the induced gap provides a guide for designing material combinations in experiments.
Although electronic correlations are weak in the thin-film semiconductor,
studying the instability of the flat-band MBSs against interactions would be an intriguing future work~\cite{sigrist_00,lee_14,schnyder_16}.

In summary, we consider a three-dimensional Josephson junction consisting of a semiconductor hosting a PSH state and a conventional $s$-wave superconductor
(see Fig.~\ref{fig:figure1}).
We demonstrate the emergence of the flat-band MBSs due to the effective $p_x$-wave superconductivity in the vicinity of the semiconductor segment.
The proposed system has three significant advantages as follows: (i) the system can be fabricated by interfacing existing materials,
(ii) the system can enter the topologically nontrivial phase without large Zeeman potentials exceeding the pair potentials,
and (iii) the system has a symmetric structure leading to a vanishing Rashba SOC potential.
Our proposal therefore represents a promising approach to realize the effective $p_x$-wave superconductor hosting the flat-band MBSs.

\begin{acknowledgments}
We are grateful to S. Tamura for the fruitful discussions.
S.I. is supported by a Grant-in-Aid for JSPS Fellows (JSPS KAKENHI Grant No. JP21J00041).
Y. T. acknowledges support from a Grant-in-Aid for Scientific Research B (KAKENHI Grant No. JP18H01176 and No. JP20H01857)
and for Scientific Research A (KAKENHI Grant No. JP20H00131), and Japan RFBR Bilateral Join Research Projects/Seminars number 19-52-50026.
This work is also supported by the JSPS Core-to-Core Program (No. JPJSCCA20170002).

D.O. and S.I. contributed equally to this work.
\end{acknowledgments}

\clearpage
\onecolumngrid
\begin{center}
 \textbf{\large Supplements Materials\\``Flat-band Majorana states in topological Josephson junctions''}\\ \vspace{0.3cm}
Daisuke Oshima$^{1}$, Satoshi Ikegaya$^{1}$, Andreas P. Schnyder$^{2}$, and Yukio Tanaka$^{1}$\\ \vspace{0.1cm}
{\itshape $^{1}$Department of Applied Physics, Nagoya University, Nagoya 464-8603, Japan\\
$^{2}$Max-Planck-Institut f\"ur Festk\"orperforschung, Heisenbergstrasse 1, D-70569 Stuttgart, Germany}
\date{\today}
\end{center}

\vspace{25pt}

\section{Detail of the derivation of the effective Hamiltonian}

In this section, we derive an effective Hamiltonian describing the subgap states localized around the Josephson junction interface.
Let us consider  a Bogoliubov--de Gennes (BdG) Hamiltonian in Eq.~(1) of the main text:
\begin{gather}
\check{H}(\boldsymbol{r}) = \check{H}_0(\boldsymbol{r}) + \check{H}_{\lambda}(\boldsymbol{r}),\\
\check{H}_0(\boldsymbol{r}) = \left[ \begin{array}{cccc}
\xi(\boldsymbol{r}) & V & 0 & \Delta(z) \\
V & \xi(\boldsymbol{r}) & -\Delta(z) & 0 \\ 
0 & -\Delta^{\ast}(z) & -\xi(\boldsymbol{r}) & -V \\
\Delta^{\ast}(z) & 0 & -V & -\xi(\boldsymbol{r}) \\
\end{array} \right],\\
\check{H}_{\lambda}(\boldsymbol{r}) = \left[ \begin{array}{cccc}
-i\lambda\delta(z)\partial_x & 0 & 0 & 0 \\
0 & +i\lambda\delta(z)\partial_x & 0 & 0 \\ 
0 & 0 & -i\lambda\delta(z)\partial_x & 0 \\
0 & 0 & 0 & +i\lambda\delta(z)\partial_x \\
\end{array} \right],\\
\xi(\boldsymbol{r}) = -\frac{\hbar^2}{2m} (\partial^2_x+\partial^2_y+\partial^2_z)-\mu,\\
\Delta(z) = \left\{ \begin{array}{cl}
\Delta e^{i\varphi/2} & \text{for } z \leq 0\\
\Delta e^{-i\varphi/2} & \text{for } z > 0
\end{array} \right.,
\end{gather}
where $m$, $\mu$, $V$, and $\lambda$ represent the effective math of an electron, chemical potential, Zeeman potential,
and the strength of the spin-orbit coupling potential sustaining a persistent spin helix state, respectively.
To derive the effective Hamiltonian for the subgap states, we treat $\check{H}_{\lambda}$ as a perturbation.
Within the zeroth-order perturbation theory, the energy eigenvalue and wave function of the subgap state satisfy the BdG equation,
\begin{gather}
\check{H}_0(\boldsymbol{r}) \check{U}_{\mathrm{SG}}(\boldsymbol{r})
= \check{U}_{\mathrm{SG}}(\boldsymbol{r}) \check{E}_0,\\
\check{E}_0=\mathrm{diag}[\varepsilon-V, \varepsilon+V, -\varepsilon+V, -\varepsilon-V],
\end{gather}
under the conditions,
\begin{align}
0 \leq \varepsilon < \Delta
\end{align}
and
\begin{align}
\lim_{z\rightarrow\pm \infty}\check{U}_{\mathrm{SG}}(\boldsymbol{r}) = 0.
\end{align}
As a preliminary step of obtaining $\varepsilon$ and $\check{U}_{\mathrm{SG}}(\boldsymbol{r})$, we first solve the BdG equation,
\begin{align}
\left[ \begin{array}{cc} \xi(\boldsymbol{r}) & s\Delta(z) \\ s\Delta^{\ast}(z) & -\xi(\boldsymbol{r}) \end{array}\right]
\left[\begin{array}{c} u_s(\boldsymbol{r}) \\ v_s(\boldsymbol{r}) \end{array}\right]
= \varepsilon \left[\begin{array}{c} u_s(\boldsymbol{r}) \\ v_s(\boldsymbol{r}) \end{array}\right],
\end{align}
for $s=\pm$.
The localized wave function for $z \leq 0$ and that for $z >0$ are represented as
\begin{align}
\psi_{b,s}(\boldsymbol{r}) = \left[\begin{array}{c} u_{b,s}(\boldsymbol{r}) \\ v_{b,s}(\boldsymbol{r}) \end{array}\right]=
\left(\frac{A_{b,s}}{\sqrt{2}} \left[\begin{array}{c} e^{i\eta} \\ s e^{-i\varphi/2} \end{array}\right] e^{-ik_+z}
+ \frac{B_{b,s}}{\sqrt{2}} \left[\begin{array}{c} s e^{i\varphi/2} \\ e^{i\eta} \end{array}\right] e^{ik_-z} \right)
\frac{e^{ik_x x}}{\sqrt{L_x}}\frac{e^{ik_y y}}{\sqrt{L_y}},
\end{align}
and
\begin{align}
\psi_{t,s}(\boldsymbol{r}) = \left[\begin{array}{c} u_{t,s}(\boldsymbol{r}) \\ v_{t,s}(\boldsymbol{r}) \end{array}\right]=
\left(\frac{A_{t,s}}{\sqrt{2}} \left[\begin{array}{c} e^{i\eta} \\ s e^{i\varphi/2} \end{array}\right] e^{ik_+z}
+ \frac{B_{t,s}}{\sqrt{2}} \left[\begin{array}{c} s e^{-i\varphi/2} \\ e^{i\eta} \end{array}\right] e^{-ik_-z} \right)
\frac{e^{ik_x x}}{\sqrt{L_x}}\frac{e^{ik_y y}}{\sqrt{L_y}},
\end{align}
respectively, where
\begin{gather}
k_{\pm} = \sqrt{\frac{2m}{\hbar^2} \sqrt{\mu^2_z+\Omega^2}} e^{\pm iQ}, \quad
2Q = \mathrm{arctan}\frac{\Omega}{\mu_z}, \quad
\mu_z = \mu - \frac{\hbar^2}{2m}(k_x^2+k_y^2),\\
\eta = \mathrm{arctan}\frac{\Omega}{\varepsilon}, \quad
\Omega = \sqrt{\Delta^2-\varepsilon^2},
\end{gather}
and we apply a periodic boundary condition in the $x-$ ($y$-) direction with $L_{x(y)}$.
Here,  $L_{x(y)}$ is the length in the $x$- ($y$-) direction.
From the boundary condition at the junction interface,
\begin{gather}
\lim_{z \rightarrow 0}\psi_{b,s}(\boldsymbol{r})  = \lim_{z \rightarrow 0} \psi_{t,s}(\boldsymbol{r}),\\
\partial_z \psi_{b,s}(\boldsymbol{r}) |_{z=0} = \partial_z \psi_{t,s}(\boldsymbol{r}) |_{z=0},
\end{gather}
and a normalization condition,
\begin{align}
\int d x \int d y \left\{ \int^{0}_{-\infty} dz \; \psi^{\dagger}_{b,s}(\boldsymbol{r}) \psi_{b,s}(\boldsymbol{r})
+ \int^{\infty}_{0} dz \; \psi^{\dagger}_{t,s}(\boldsymbol{r}) \psi_{t,s}(\boldsymbol{r}) \right\}=1,
\end{align}
we obtain the energy eigenvalue as
\begin{gather}
\varepsilon = \Delta \sqrt{1-D \sin^2 (\varphi/2)},\\
D = \frac{\Delta^2\sin^2 (\varphi/2)-\mu_z^2+\mu_z \sqrt{2\Delta^2\sin^2 (\varphi/2)+\mu_z^2}}{2\Delta^2\sin^2 (\varphi/2)},
\end{gather}
and the numerical coefficients in the wave function as
\begin{gather}
A_{b,s}=\frac{-s i e^{-iQ/2} f_- }{\sqrt{N}}, \quad
B_{b,s}=\frac{e^{-i \varphi/2} e^{iQ/2} f_+ }{\sqrt{N}}, \quad
A_{t,s}=\frac{s e^{-i \varphi/2} e^{-iQ/2} f_+ }{\sqrt{N}}, \quad
B_{t,s}=\frac{ i e^{iQ/2} f_- }{\sqrt{N}},\\
f_{\pm} = \sqrt{\frac{1}{2}\left( 1 \pm \frac{\sqrt{D \sin^2 (\varphi/2)}}{\sqrt{1-D \sin^2 (\varphi/2)}}\cot (\varphi/2)\right)},\quad
N = \frac{1}{\kappa}D (3-2D), \quad
\kappa = \sqrt{\frac{2m}{\hbar^2} \sqrt{\mu^2_z+\Omega^2}} \sqrt{1-D},
\end{gather}
where the condition of
\begin{align}
-\frac{1}{2}\Delta|\sin(\varphi/2)| < \mu_z,
\end{align}
must be satisfied to realize the bound states, (i.e., $\lim_{z\rightarrow\pm \infty}\psi_{t,s}(\boldsymbol{r})=0$).
From particle-hole symmetry of the superconducting system, we also obtain
\begin{align}
\left[ \begin{array}{cc} \xi(\boldsymbol{r}) & s\Delta(z) \\ s\Delta^{\ast}(z) & -\xi(\boldsymbol{r}) \end{array}\right]
\left[\begin{array}{c} -v^{\ast}_s(\boldsymbol{r}) \\ u_s^{\ast}(\boldsymbol{r}) \end{array}\right]
= -\varepsilon \left[\begin{array}{c} -v^{\ast}_s(\boldsymbol{r}) \\ u_s^{\ast}(\boldsymbol{r}) \end{array}\right].
\end{align}
Finally, we obtain
\begin{gather}
\check{U}_{\mathrm{SG}}(\boldsymbol{r}) = \check{Q} \; \check{U}^{\prime}_{\mathrm{SG}}(\boldsymbol{r}) \; \check{W},\\
\check{U}^{\prime}_{\mathrm{SG}}(\boldsymbol{r})=
\left[ \begin{array}{cccc}
u_+(\boldsymbol{r}) & 0 & 0 & -v_+^{\ast}(\boldsymbol{r}) \\
0 & u_-(\boldsymbol{r}) & -v_-^{\ast}(\boldsymbol{r}) & 0 \\ 
0 & v_-(\boldsymbol{r}) & u_-^{\ast}(\boldsymbol{r}) & 0 \\
v_+(\boldsymbol{r}) & 0 & 0 & u_+^{\ast}(\boldsymbol{r}) \\
\end{array} \right],\\
\check{Q}=\left[ \begin{array}{cc} \hat{Q} & 0 \\ 0 & \hat{Q} \end{array} \right], \quad
\hat{Q}=\frac{1}{\sqrt{2}}\left[ \begin{array}{cc} 1 & 1 \\ -1 & 1 \end{array} \right],\quad
\check{W}=\left[ \begin{array}{cc} \hat{w} & 0 \\ 0 & \hat{w}^{\ast} \end{array} \right], \quad
\hat{w}=\left[ \begin{array}{cc} i e^{-i\eta/2}e^{i\varphi/4} & 0 \\ 0 & -i e^{-i\eta/2}e^{i\varphi/4} \end{array} \right],
\end{gather}
where we introduce $\check{W}$ for later convenience.
From the first-order perturbation theory, we construct the effective Hamiltonian for the subgap states as,
\begin{align}
\check{H}_{\mathrm{eff}} &= \int d \boldsymbol{r}
\check{U}^{\dagger}_{\mathrm{SG}}(\boldsymbol{r}) \check{H}(\boldsymbol{r}) \check{U}_{\mathrm{SG}}(\boldsymbol{r}) \nonumber\\
&=\int d \boldsymbol{r} \check{U}^{\dagger}_{\mathrm{SG}}(\boldsymbol{r}) \check{H}_0(\boldsymbol{r}) \check{U}_{\mathrm{SG}}(\boldsymbol{r})
+ \int d \boldsymbol{r} \check{U}^{\dagger}_{\mathrm{SG}}(\boldsymbol{r}) \check{H}_{\lambda}(\boldsymbol{r}) \check{U}_{\mathrm{SG}}(\boldsymbol{r})
\nonumber\\
&=\left[ \begin{array}{cccc}
\varepsilon -V & \tilde{\lambda}k_x & \tilde{\Delta}k_x & 0 \\
\tilde{\lambda}k_x & \varepsilon +V & 0 & -\tilde{\Delta}k_x \\ 
\tilde{\Delta}k_x & 0 & -\varepsilon +V & \tilde{\lambda}k_x \\
0 & -\tilde{\Delta}k_x & \tilde{\lambda}k_x & -\varepsilon -V \\
\end{array} \right],
\end{align}
with
\begin{gather}
\tilde{\lambda} = \lambda \sqrt{D(1-D)}F,\\
\tilde{\Delta} = \lambda \cot (\varphi/2) F,\\
F = \sqrt{\frac{2m}{\hbar^2} \sqrt{\mu^2_z+\Omega^2}} \frac{\sqrt{1-D}}{D(3-2D)}\frac{\sqrt{D \sin^2 (\varphi/2)}}{\sqrt{1-D \sin^2 (\varphi/2)}},
\end{gather}
where $\tilde{\lambda}$ and $\tilde{\Delta}$ are the even-functions with respect to $k_x$ and $k_y$.
As also discussed in the main text, we find that
the effective Hamiltonian of $\check{H}_{\mathrm{eff}}$ is equivalent to the BdG Hamiltonian of a two-dimensional $p_x$-wave superconductor
in the presence of the spin-orbit coupling potential and the Zeeman potential.

Then, we analyze the topological property of $\check{H}_{\mathrm{eff}}$.
The energy eigenvalues of $\check{H}_{\mathrm{eff}}$ are given by
\begin{align}
E_{s}=\pm \sqrt{\left( \varepsilon + s \sqrt{V^2+\tilde{\lambda}^2k_x^2}\right)^2 + \tilde{\Delta}^2 k_x^2},
\end{align}
for $s=\pm$.
While $E_+$ is fully gapped irrespective of $\boldsymbol{k}_{\parallel}=(k_x,k_y)$,
the branch of $E_-$ can have superconducting gap nodes when
\begin{gather}
\varepsilon^2 = V^2 + \tilde{\lambda}^2 k_x^2,\\
\tilde{\Delta} k_x = 0,
\end{gather}
is satisfied.
At $k_x=0$, this condition is simplified to
\begin{align}
\varepsilon^2 = V^2.
\label{eq:node}
\end{align}
From Eq.~(\ref{eq:node}), we find the two superconducting gap nodes when the condition
\begin{gather}
X < V^2 <\Delta^2,
\label{eq:condition}\\
X = \frac{\mu^2+\Delta^2 \left\{1+\cos^2 (\varphi/2)\right\}-\mu \sqrt{\mu^2+2\Delta^2\sin^2 (\varphi/2)}}{2},
\end{gather}
is satisfied; the two gap nodes are located at
\begin{gather}
\boldsymbol{k}_{\parallel}=(0,\pm \tilde{k}_F),\\
\tilde{k}_F=k_F\sqrt{1+\zeta},\\
\zeta = \frac{1}{\mu}\left[V^2-\frac{\Delta^2}{2}\left\{1+\cos^2 (\varphi/2)\right\}\right]\sqrt{\frac{1}{V^2-\Delta^2\cos^2 (\varphi/2)}},
\end{gather}
with $k_F=\sqrt{2m\mu}/\hbar$.
In the limit of $\mu \gg \Delta$, this condition in Eq.~(\ref{eq:condition}) is reduced to
\begin{align}
\Delta^2\cos^2 (\varphi/2)<V^2<\Delta^2.
\end{align}
In addition, $E_-$ at $\varphi=\pi$ becomes gapless irrespective of $\boldsymbol{k}_{\parallel}$
because the effective pair potential $\tilde{\Delta} \propto \cot(\phi/2)$ vanishes.
We have two distinct nodal phases separated by $\varphi=\pi$:
the nodal phase I for $\varphi<\pi$ and the nodal phase II for $\varphi>\pi$, as shown in Fig.~2(a) of the main text.
The effective Hamiltonian of $\check{H}_{\mathrm{eff}}$ has chiral symmetry defined by
\begin{gather}
\check{S}_{\mathrm{eff}} \check{H}_{\mathrm{eff}} (\boldsymbol{k}_{\parallel}) \check{S}_{\mathrm{eff}}^{-1}
=-\check{H}_{\mathrm{eff}} (\boldsymbol{k}_{\parallel}),\quad
\check{S}_{\mathrm{eff}} = \left[ \begin{array}{cccc}
0 & 0 & -i & 0 \\
0 & 0 & 0 & i \\ 
i & 0 & 0 & 0 \\
0 & -i & 0 & 0 \\
\end{array} \right],
\end{gather}
where the original BdG Hamiltonian $\check{H}(\boldsymbol{r})$ also preserves chiral symmetry as
\begin{gather}
\check{S} H (\boldsymbol{r}) \check{S}^{-1}=-H (\boldsymbol{r}),\\
\check{S}=\check{M}_{xy}\check{T}\check{C},\\
\check{M}_{xy}=\left[ \begin{array}{cccc}
-i & 0 & 0 & 0 \\
0 & i & 0 & 0 \\ 
0 & 0 & i & 0 \\
0 & 0 & 0 & -i \\
\end{array} \right]\mathcal{R}_z,\quad
\check{T}=\left[ \begin{array}{cccc}
0 & 1 & 0 & 0 \\
-1 & 0 & 0 & 0 \\ 
0 & 0 & 0 & 1 \\
0 & 0 & -1 & 0 \\
\end{array} \right] \mathcal{K},\quad
\check{C}=\left[ \begin{array}{cccc}
0 & 0 & 1 & 0 \\
0 & 0 & 0 & 1 \\ 
1 & 0 & 0 & 0 \\
0 & 1 & 0 & 0 \\
\end{array} \right] \mathcal{K},
\end{gather}
where $\check{M}_{xy}$, $\check{T}$, and $\check{C}$ represent
mirror reflection symmetry with respect to the $xy$-plane, time-reversal symmetry, and particle-hole symmetry, respectively;
$\mathcal{R}_z$ describes the reflection of spatial coordinate z (i.e., $z\rightarrow-z$), and  $\mathcal{K}$ is the complex-conjugation operator.
We can characterize the topological property of $\check{H}_{\rm eff}$ by using a one-dimensional winding number~\cite{sato_11},
\begin{align}
w(k_y) = \frac{i}{4 \pi}
\int dk_x \mathrm{Tr} [ \check{S}_{\mathrm{eff}} \left\{ \check{H}_{\mathrm{eff}}(\boldsymbol{k}_{\parallel}) \right\}^{-1}
\partial_{k_x} \check{H}_{\mathrm{eff}}(\boldsymbol{k}_{\parallel}) ].
\end{align}
It is possible to compute the winding number more simply by employing the following procedures.
First, according to Eq.~(\ref{eq:node}), the emergence/position of the nodes is irrelevant to $\tilde{\lambda}$.
Therefore, since the topological number is unchanged except when the superconducting gap closes,
we can evaluate the winding number precisely with setting $\tilde{\lambda}\rightarrow 0$.
In this limit, $\check{H}_{\mathrm{eff}}$ can be block diagonalized as
$\check{H}_{\mathrm{eff}}\rightarrow \mathrm{diag}[\hat{H}_+, \hat{H}_-]$, where
\begin{align}
\hat{H}_s=\left[ \begin{array}{cc} \varepsilon +sV & -s \tilde{\Delta}k_x \\ -s \tilde{\Delta}k_x & -\varepsilon-sV \end{array} \right]
\end{align}
for $s=\pm$.
The chiral symmetry of each block component is given by
\begin{align}
\hat{S}_s \hat{H}_s (\boldsymbol{k}_{\parallel}) \hat{S}_s^{-1}=-\hat{H}_s (\boldsymbol{k}_{\parallel}), \quad
\hat{S}_s=s\left[ \begin{array}{cc} 0 & i \\ -i & 0 \end{array} \right].
\end{align}
Since the superconducting gap nodes are originated from the $\hat{H}_-$ component, we can assess the winding number solely from $\hat{H}_-$.
As a result, the winding number can be calculated by
\begin{align}
w(k_y) &= \frac{i}{4 \pi}
\int dk_x \mathrm{Tr} [ \hat{S}_- \left\{ \hat{H}_-(\boldsymbol{k}_{\parallel}) \right\}^{-1}
\partial_{k_x} \hat{H}_- (\boldsymbol{k}_{\parallel}) ] . \label{eq:wind}
\end{align}
According to the strict definition for the winding number in continuum model,
the integration in Eq.~(\ref{eq:wind}) is performed from $k_x=-\infty$ to $\infty$,
where we need to apply an additional hypothetical regulation, i.e., $\tilde{\Delta} \rightarrow 0$ at $k_x=\pm \infty$,
so that we obtain $\hat{H}_-(k_x=\infty)=\hat{H}_-(k_x=-\infty)$~\cite{sato_11}.
Even so, we can practically compute the winding number by using a simplified formula,
which is obtained independently of the details of the treatment at $k_x=\pm\infty$~\cite{sato_11}:
\begin{align}
w(k_y) =\frac{1}{2} \sum_{\varepsilon(\boldsymbol{k}_{\parallel}) =V}
\mathrm{sgn} [ \tilde{\Delta}k_x ]
\mathrm{sgn} [ \partial_{k_x} \varepsilon (\boldsymbol{k}_{\parallel}) ],
\end{align}
where $\sum_{\varepsilon(\boldsymbol{k}_{\parallel})=V}$ represents the summation over $k_x$
satisfying $\varepsilon(\boldsymbol{k}_{\parallel})=V$ with the fixed $k_y$.
Finally, we obtain
\begin{align}
w(k_y)=
\left\{ \begin{array}{cl} 
+1& \text{for}\quad |k_y|< \tilde{k}_F \\
0 & \text{otherwise}
\end{array}\right.,
\label{eq: wind_I}
\end{align}
for the nodal phase I in Fig. 2(a) of the main text, and
\begin{align}
w(k_y)=
\left\{ \begin{array}{cl} 
-1& \text{for}\quad |k_y|< \tilde{k}_F \\
0 & \text{otherwise}
\end{array}\right.,
\label{eq: wind_II}
\end{align}
for the nodal phase II in Fig. 2(a) of the main text, while $w(k_y)=0$ irrespective of $k_y$ for the fully gapped phase.

\section{Flat-band Majorana bound states in the presence of perturbations}
In this section, we show $\rho_{\mathrm{edge}}(k_y;E)$ in the presence of the perturbations discussed in the main text.
Hereafter, the results of $\rho_{\mathrm{edge}}(k_y;E)$ are normalized by the same unit as the one in Fig. 2(c) of the main text,
and we plot $\log_{10}\rho_{\mathrm{edge}}(k_y;E)$ instead of the raw data of $\rho_{\mathrm{edge}}(k_y;E)$.
In addition, we show $\bar{\rho}_{\mathrm{edge}}(E=0)$ normalized by the same unit as the one in Fig. 2(b) of the main text.

\begin{figure}[hhhh]
\begin{center}
\includegraphics[width=0.98\textwidth]{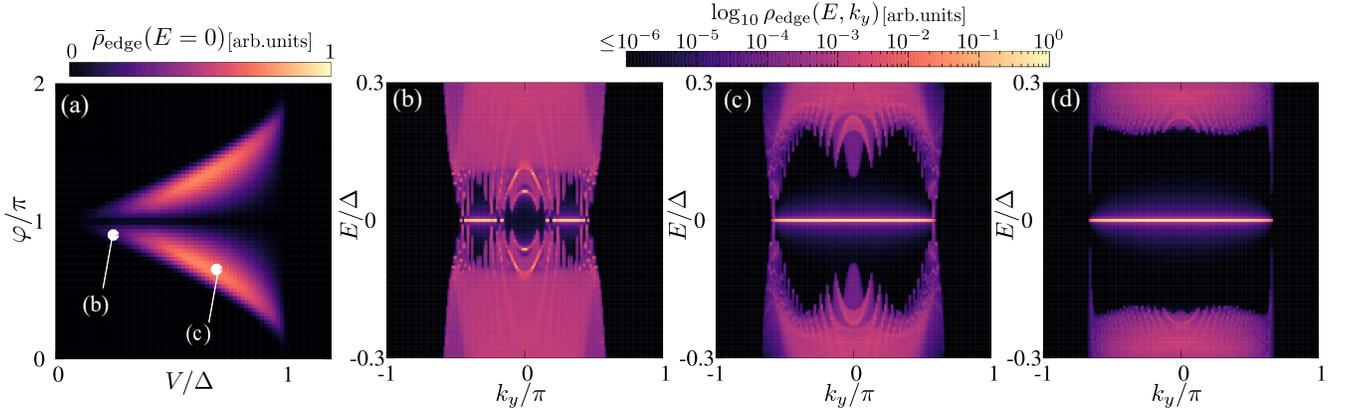}
\caption{(a) Zero-energy LDOS,  $\bar{\rho}_{\rm edge}(E=0)$, as a function of the Zeeman potential $V$ and the superconducting phase difference $\varphi$,
where we choose $t_1=t_2=-2\Delta$; the system preserves both chiral symmetry and modified particle-hole symmetry.
This figure is equivalent to Fig.~3(a) in the main text.
In (b) and (c), we show $\rho_{\rm edge}(k_y;E)$ on a log-scale, i.e., $\log_{10} \rho_{\rm edge}(k_y;E)$, as a function of the energy $E$ and the momentum $k_y$,
where we choose (b) $(V,\varphi)=(0.25\Delta,0.9\pi)$ and (c) $(V,\varphi)=(0.7\Delta,0.65\pi)$, respectively.
In (d), for comparison, we plot $\log_{10} \rho_{\rm edge}(k_y;E)$ for the ideal junction (i.e., $t_1=t_2=0$), where we choose $(V,\varphi)=(0.7\Delta,0.65\pi)$.}
\label{fig:figure_sup1}
\end{center}
\end{figure}

\begin{figure}[hhhh]
\begin{center}
\includegraphics[width=0.6\textwidth]{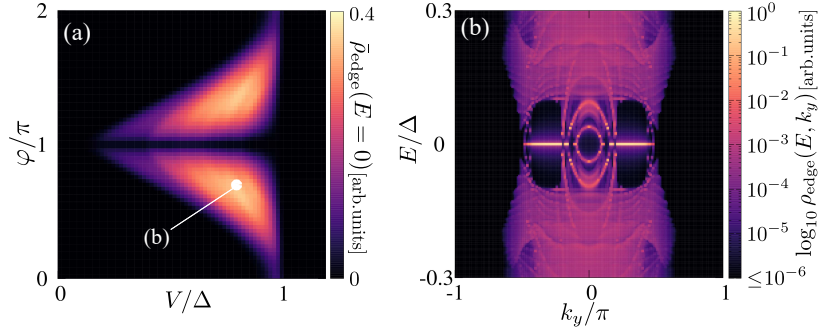}
\caption{(a) Zero-energy LDOS,  $\bar{\rho}_{\rm edge}(E=0)$, as a function of $V$ and $\varphi$,
where the system preserves both chiral symmetry and modified particle-hole symmetry with $t_1=t_2=-5\Delta$.
In (b), we show $\log_{10} \rho_{\rm edge}(k_y;E)$ as a function of $E$ and $k_y$, where we choose $(V,\varphi)=(0.8\Delta,0.7\pi)$.}
\label{fig:figure_sup2}
\end{center}
\end{figure}

Firstly, we consider the system preserving both chiral symmetry [see Eq.~(12) of the main text]
and modified particle-hole symmetry [see Eq.~(18) of the main text],
where the junction transparency is decreased by the perturbation $H_{\delta t}$ with $t_1=t_2$ [see Eq~(28) of the main text].
Figure~\ref{fig:figure_sup1}(a) shows the $\bar{\rho}_{\mathrm{edge}}(E=0)$ with $t_1=t_2=-2\Delta$ as a function of $V$ and $\varphi$,
which is equivalent to Fig.~3(a) in the main text.
In Figs.~\ref{fig:figure_sup1}(b) and \ref{fig:figure_sup1}(c), we show $\rho_{\rm edge}(k_y;E)$ as a function of $k_y$ and $E$,
where we choose (b) $(V,\varphi)=(0.25\Delta,0.9\pi)$ and (c) $(V,\varphi)=(0.7\Delta,0.65\pi)$, respectively [see also the dots in Figs.~\ref{fig:figure_sup1}(a) ].
Since both chiral symmetry and modified particle-hole symmetry are preserved, we can find the flat-band MBSs.
Nevertheless, in the vicinity of the topological phase boundary, as shown in Fig.~\ref{fig:figure_sup1}(b),
the flat-band MBSs appear at several separated regions with respect to $k_y$, where we also find more than two gap nodes;
similar flat-band MBSs also appear in the other effective $p_x$-wave superconductors~\cite{ikegaya_21,ikegaya_18,law_13,rosenow_14}.
In Fig.~\ref{fig:figure_sup1}(d), we plot $\log_{10} \rho_{\rm edge}(k_y;E)$ for the ideal junction (i.e., $t_1=t_2=0$),
where we choose $(V,\varphi)$ as same as that in Fig.~\ref{fig:figure_sup1}(c) [i.e.,  $(V,\varphi)=(0.7\Delta,0.65\pi)$].
From the comparison between Fig.~\ref{fig:figure_sup1}(c) and Fig.~\ref{fig:figure_sup1}(d), especially around $k_y=0$,
we see that the induced gap is suppressed by decreasing the junction transparency.
In Figs.~\ref{fig:figure_sup2}(a) and \ref{fig:figure_sup2}(b), we consider the junction with decreasing the transparency further, i.e., $t_1=t_2=-5\Delta$.
In Fig.~\ref{fig:figure_sup2}(a), we plot $\bar{\rho}_{\mathrm{edge}}(E=0)$ in the range from $0$ to $0.4$,
while that in Fig.~\ref{fig:figure_sup1}(a) is shown in the range from $0$ to $1$ [see the color bars in Figs.~\ref{fig:figure_sup1}(a) and \ref{fig:figure_sup2}(a)].
We see that $\bar{\rho}_{\mathrm{edge}}(E=0)$ is suppressed by decreasing the junction transparency.
Figure~\ref{fig:figure_sup2}(b) shows $\rho_{\rm edge}(k_y;E)$ at $(V,\varphi)=(0.8\Delta,0.7\pi)$,
which is located at deep inside of the topologically nontrivial region.
We see that the suppression of the induced gap becomes more evident by decreasing the junction transparency.
Especially, the induced gap around $k_y=0$ is disturbed devastatingly, and thus the flat-band MBSs around $k_y=0$ disappear.
We note that the reduction of the induced gap by decreasing the junction transparency is also found in the planar TJJs~\cite{halperin_17}.

\begin{figure}[hhhh]
\begin{center}
\includegraphics[width=0.75\textwidth]{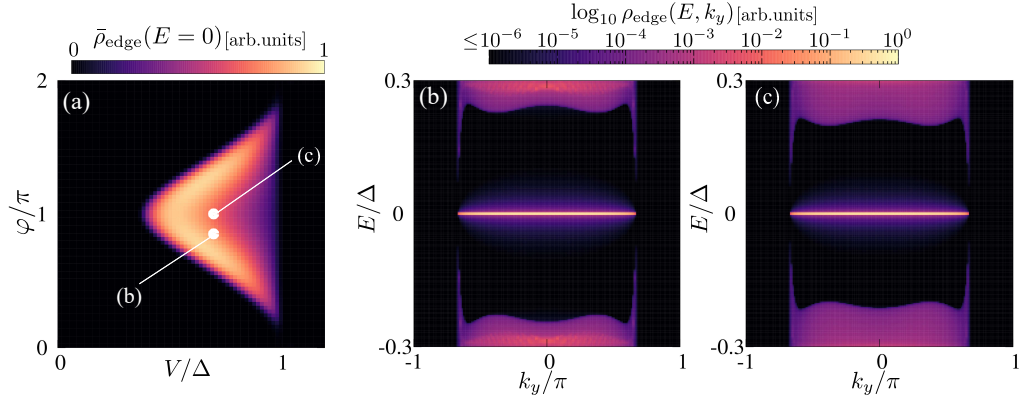}
\caption{(a) Zero-energy LDOS,  $\bar{\rho}_{\rm edge}(E=0)$, as a function of $V$ and $\varphi$,
where we consider the asymmetric junction transparency with $t_1=2\Delta$ and $t_2=-2\Delta$;
the chiral symmetry is broken, while the modified particle-hole symmetry is still preserved.
This figure is equivalent to Fig.~3(b) in the main text.
In (b) and (c), we show $\log_{10} \rho_{\rm edge}(k_y;E)$ as a function of $E$ and $k_y$,
where we choose (b) $(V,\varphi)=(0.7\Delta,0.85\pi)$ and (c) $(V,\varphi)=(0.7\Delta,\pi)$, respectively.}
\label{fig:figure_sup3}
\end{center}
\end{figure}

Secondary, we consider the system with broken chiral symmetry by $t_1\neq t_2$, where the modified particle-hole symmetry is still preserved.
Figure~\ref{fig:figure_sup3}(a) shows the $\bar{\rho}_{\mathrm{edge}}(E=0)$ with $t_1=2\Delta$ and $t_2=-2\Delta$, which is equivalent to Fig.~3(b) in the main text.
In Figs.~\ref{fig:figure_sup3}(b) and \ref{fig:figure_sup3}(c), we show $\rho_{\rm edge}(k_y;E)$,
where we choose (b) $(V,\varphi)=(0.7\Delta,0.85\pi)$ and (c) $(V,\varphi)=(0.7\Delta,\pi)$, respectively.
Although the chiral symmetry is broken, since the modified particle-hole symmetry is still preserved,
we clearly find the emergence of the flat-band MBSs protected by the $\mathbb{Z}_2$ index.
Moreover, as shown in Figs.~\ref{fig:figure_sup3}(c), we obtain the flat-band MBSs at $\varphi=\pi$.
The results imply that the two nodal phases originally separated by $\varphi=\pi$ merge into a single nodal phase hosting the flat-band MBSs.

\begin{figure}[hhhh]
\begin{center}
\includegraphics[width=0.98\textwidth]{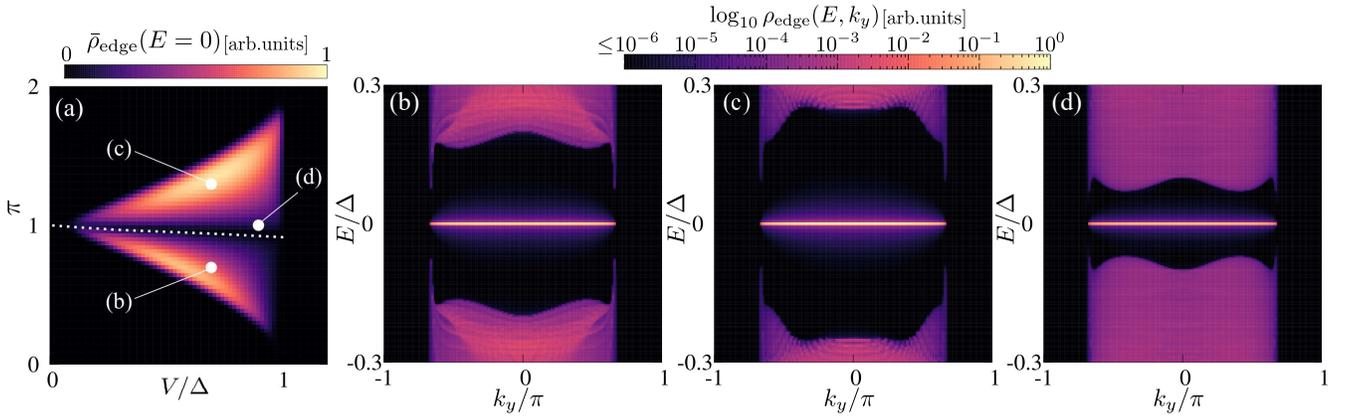}
\caption{(a) Zero-energy LDOS,  $\bar{\rho}_{\rm edge}(E=0)$, as a function of $V$ and $\varphi$,
where we consider the Rashba SOC potential with $\lambda^\prime_1=-\lambda^\prime_2=2.5\Delta$;
the modified particle-hole symmetry is broken, while the chiral symmetry is still preserved.
The dotted line indicates $(V,\varphi)$ taking the local minima of $\bar{\rho}_{\rm edge}(E=0)$.
This figure is equivalent to Fig.~3(c) in the main text.
In (b)-(d), we show $\log_{10} \rho_{\rm edge}(k_y;E)$ as a function of $E$ and $k_y$,
where we choose (b) $(V,\varphi)=(0.7\Delta,0.7\pi)$, (c) $(V,\varphi)=(0.7\Delta,1.3\pi)$, and (d) $(V,\varphi)=(0.9\Delta,\pi)$, respectively.}
\label{fig:figure_sup4}
\end{center}
\end{figure}

\begin{figure}[hhhh]
\begin{center}
\includegraphics[width=0.75\textwidth]{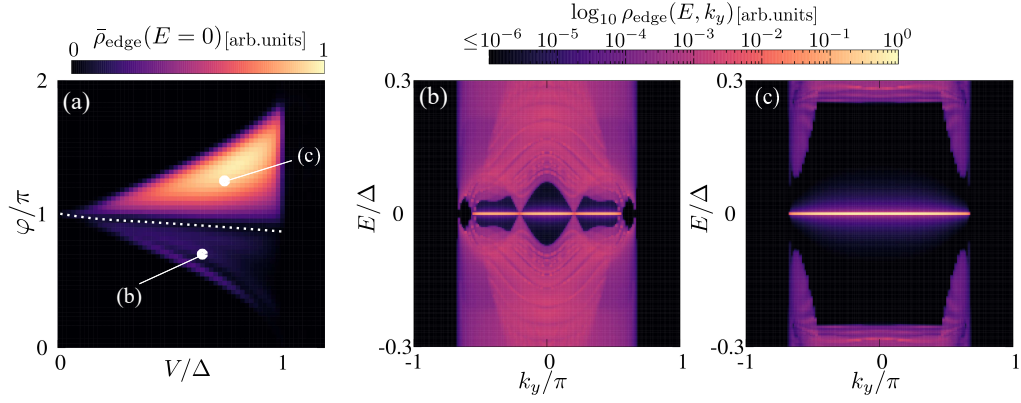}
\caption{(a) Zero-energy LDOS,  $\bar{\rho}_{\rm edge}(E=0)$, as a function of $V$ and $\varphi$,
where we choose $\lambda^\prime_1=-\lambda^\prime_2=5\Delta$;
the modified particle-hole symmetry is broken, while the chiral symmetry is still preserved.
The dotted line indicates $(V,\varphi)$ taking the local minima of $\bar{\rho}_{\rm edge}(E=0)$.
In (b) and (c), we show $\log_{10} \rho_{\rm edge}(k_y;E)$ as a function of $E$ and $k_y$,
where we choose (b) $(V,\varphi)=(0.65\Delta,0.7\pi)$ and (c) $(V,\varphi)=(0.75\Delta,1.25\pi)$, respectively.}
\label{fig:figure_sup5}
\end{center}
\end{figure}

Thirdly, we consider the system in the presence of the Rashbe spin-orbit coupling (SOC) potential [see Eq.~(29) of the main text]
with preserving the chiral symmetry (i.e., $\lambda^\prime_1=-\lambda^\prime_2$), while the modified particle-hole symmetry is broken.
Figure~\ref{fig:figure_sup4}(a) shows the $\bar{\rho}_{\mathrm{edge}}(E=0)$ with $\lambda^\prime_1=-\lambda^\prime_2=2.5\Delta$,
which is equivalent to Fig.~3(c) in the main text.
In Figs.~\ref{fig:figure_sup4}(b) and \ref{fig:figure_sup4}(c), we show $\rho_{\rm edge}(k_y;E)$ at (b) $(V,\varphi)=(0.7\Delta,0.7\pi)$
and (c) $(V,\varphi)=(0.7\Delta,1.3\pi)$, respectively.
Since the chiral symmetry is preserved, we obtain the flat-band MBSs.
We additionally find that the induced gaps become asymmetric with respect to $\varphi = \pi$, i.e.,
the induced gap for $\varphi<\pi$ [Fig.~\ref{fig:figure_sup4}(b)] is slightly smaller than that for $\varphi>\pi$ [Fig.~\ref{fig:figure_sup4}(c)].
Such tendency becomes more evident by increasing the Rashba SOC potential.
Figure~\ref{fig:figure_sup5}(a) shows the $\bar{\rho}_{\mathrm{edge}}(E=0)$ with $\lambda^\prime_1=-\lambda^\prime_2=5\Delta$.
In Figs.~\ref{fig:figure_sup5}(b) and \ref{fig:figure_sup5}(c), we show $\rho_{\rm edge}(k_y;E)$ at
(b) $(V,\varphi)=(0.65\Delta,0.7\pi)$ and (c) $(V,\varphi)=(0.75\Delta,1.25\pi)$, respectively.
As shown in Fig.~\ref{fig:figure_sup5}(a), $\bar{\rho}_{\mathrm{edge}}(E=0)$ for $\varphi<\pi$ is distinctly smaller than that for $\varphi>\pi$.
As shown in Figs.~\ref{fig:figure_sup5}(b) and \ref{fig:figure_sup5}(c), although we obtain the flat-band MBSs in both $\varphi<\pi$ and $\varphi>\pi$,  
the induced gap for $\varphi<\pi$ is strongly suppressed; we additionally find that the energy spectrum has more than two gap nodes.
The decay length of the flat-band MBS becomes longer as the energy gap becomes smaller,
and thus the amplitude of wave function of the flat-band MBSs at the surface (i.e., $i_x=0$) becomes smaller.
This explains the suppression of surface LDOS, $\bar{\rho}_{\rm edge}(E=0)$, for $\varphi<\pi$.
Strictly speaking, the phase boundary between two distinct nodal phases slightly deviates from $\varphi=\pi$;
see dotted lines in Fig.~\ref{fig:figure_sup4}(a) and Fig.~\ref{fig:figure_sup5}(a).
Actually, as shown in Fig.~\ref{fig:figure_sup4}(d), we confirm that the flat-band MBSs appears event at $\varphi=\pi$.

\begin{figure}[hhhh]
\begin{center}
\includegraphics[width=0.6\textwidth]{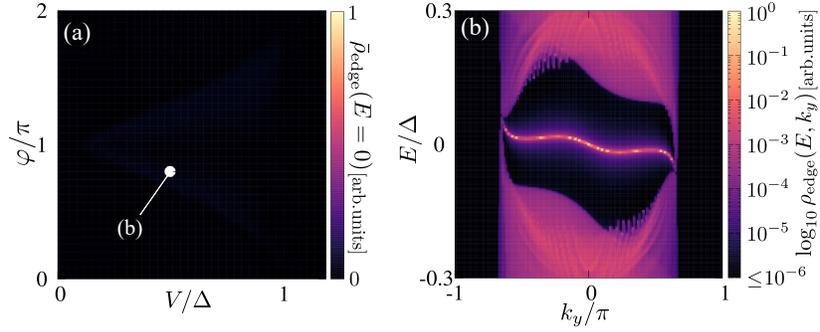}
\caption{(a) Zero-energy LDOS,  $\bar{\rho}_{\rm edge}(E=0)$, as a function of $V$ and $\varphi$,
where both chiral symmetry and modified particle-hole symmetry are broken
by the Rashba SOC potentials with $\lambda^\prime_1=3.5\Delta$ and $\lambda^\prime_2=-1.5\Delta$.
This figure is equivalent to Fig.~3(d) in the main text.
In (b), we show $\log_{10} \rho_{\rm edge}(k_y;E)$ as a function of $E$ and $k_y$, where we choose $(V,\varphi)=(0.5\Delta,0.8\pi)$.}
\label{fig:figure_sup6}
\end{center}
\end{figure}

Lastly, we consider the Rashba SOC potential breaking both chiral symmetry and modified particle-hole symmetry (i.e., $\lambda^\prime_1\neq -\lambda^\prime_2$).
Figure~\ref{fig:figure_sup6}(a) shows the $\bar{\rho}_{\mathrm{edge}}(E=0)$ with $\lambda^\prime_1=3.5\Delta$ and $\lambda^\prime_2=-1.5\Delta$,
which is equivalent to Fig.~3(d) in the main text.
The strong suppression in $\bar{\rho}_{\mathrm{edge}}(E=0)$ suggests the absence of the flat-band MBSs.
In Fig.~\ref{fig:figure_sup6}(b), we show $\rho_{\rm edge}(k_y;E)$ at $(V,\varphi)=(0.5\Delta,0.8\pi)$.
We find that the asymmetric Rashba SOC potential causes the finite slope in the dispersion of flat-band MBSs.

\clearpage

\end{document}